\documentclass[aps,prl,preprint,superscriptaddress]{revtex4-1}
\usepackage{amssymb}
\usepackage{color}
\usepackage{graphicx}
\usepackage{epsfig}
\usepackage{epstopdf}

\usepackage{amsmath}

\usepackage{notes2bib}
\bibnotesetup{
note-name = ,
use-sort-key = false,
placement = mixed
}

\newcommand{\grl}{    {Geophys. Res. Lett.}}
\newcommand{\jgr}{    {J. Geophys. Res.}}
\newcommand{\ssr}{    {Space Sci. Rev.}}
\newcommand{\planss}{    {Plan. Sp. Sci.}}

\newcommand{\blue}{\textcolor{black}}

\begin{document}


\title{Nonresonant scattering of relativistic electrons by electromagnetic ion cyclotron waves in Earth's radiation belts}


\author{Xin An}
\email[]{phyax@ucla.edu}
\affiliation{Department of Earth, Planetary, and Space Sciences, University of California, Los Angeles, CA, 90095, USA}

\author{Anton Artemyev}
\affiliation{Department of Earth, Planetary, and Space Sciences, University of California, Los Angeles, CA, 90095, USA}

\author{Vassilis Angelopoulos}
\affiliation{Department of Earth, Planetary, and Space Sciences, University of California, Los Angeles, CA, 90095, USA}

\author{Xiaojia Zhang}
\affiliation{Department of Earth, Planetary, and Space Sciences, University of California, Los Angeles, CA, 90095, USA}

\author{Didier Mourenas}
\affiliation{CEA, DAM, DIF, Arpajon, 91297, France}
\affiliation{Laboratoire Mati\`ere en Conditions Extr\^emes, Paris-Saclay University, CEA, Bruy\`eres-le-Ch\^atel, 91190, France}

\author{Jacob Bortnik}
\affiliation{Department of Atmospheric and Oceanic Sciences, University of California, Los Angeles, CA, 90095, USA}



\date{\today}

\begin{abstract}
Electromagnetic ion cyclotron waves are expected to pitch-angle scatter and cause atmospheric precipitation of relativistic ($> 1$\,MeV) electrons under typical conditions in Earth's radiation belts. However, it has been a longstanding mystery how relativistic electrons in the hundreds of keV range (but $<1$\,MeV), which are not resonant with these waves, precipitate simultaneously with those $>1$\,MeV. We demonstrate that, when the wave packets are short, nonresonant interactions enable such scattering of $100$s of keV electrons by introducing a spread in wavenumber space. We generalize the quasi-linear diffusion model to include nonresonant effects. The resultant model exhibits an exponential decay of the scattering rates extending below the minimum resonant energy depending on the shortness of the wave packets. This generalized model naturally explains observed nonresonant electron precipitation in the hundreds of keV concurrent with $>1$\,MeV precipitation.
\end{abstract}


\maketitle
The dynamics of Earth's radiation belts, and of many other space plasma systems, are largely controlled by resonant wave-particle interactions \cite{Andronov&Trakhtengerts64,Kennel&Petschek66}. Quasi-linear theory of resonant diffusion \cite{Vedenov62,Drummond&Pines62} has been the main theoretical framework for describing energetic particle scattering \cite{Shprits08:JASTP_local,Li&Hudson19,thorne2010radiation}. In quasi-linear models, diffusion rates are evaluated using statistical averages of (small amplitude) waves and background plasmas based on observations. One of the most important wave modes resulting in particle scattering and precipitation of relativistic electrons into Earth's atmosphere is the electromagnetic ion cyclotron (EMIC) mode \cite{Thorne&Kennel71,Blum15:precipitation, Usanova14, Shprits17, Kubota&Omura17, Grach&Demekhov20}. Detailed comparisons between theoretical predictions of precipitating electron energies and low-altitude precipitation measurements, however, reveal a significant discrepancy: the observed precipitation events often contain sub-MeV electrons \cite{Capannolo19:microburst,Hendry17}, at energies well below the minimum resonant energy \cite[usually $\geq 1$ MeV; see][]{Summers&Thorne03,Ni15} of EMIC waves. This discrepancy cannot be reconciled by hot plasma effects on EMIC wave dispersion \cite{cao2017scattering,Chen19}. The most promising approach is the inclusion of nonresonant electron scattering by EMIC waves \cite{Chen16:nonresonant}. In this Letter we formulate quasi-linear diffusion for nonresonant wave-particle interactions, and use it to demonstrate the impact of short (i.e., having a few wave periods in one packet) EMIC wave packets on scattering of sub-relativistic electrons. Our results naturally explain low-altitude observations of such precipitation. \blue{The effects of finite wave packets have been studied in a wide range of contexts, such as Langmuir turbulence \cite{goldman1984strong,*robinson1997nonlinear,*muschietti1994interaction,*Krafft13,*rowland1977simulations,*gurnett1976electron,*kellogg2003langmuir,*anderson1981plasma,*gurnett1981parametric,*dubois1993excitation,*leung1982plasma,*sun2022electron,*rubenchik1991strong} and time domain structures \cite{Mozer15,Vasko17:diffusion} in space and laboratory plasmas, and current drive in fusion devices \cite{fisch2003current,dodin2005nonlinear,lamb1984behavior}. The proposed approach for the inclusion of nonresonant effects into the quasi-linear diffusion formalism may be used in such plasma systems, where short wave packets are sufficiently strong to provide appreciable nonresonant particle scattering.}

To motivate the need for a generalized diffusion model including both resonant and nonresonant wave-particle interactions, we examine an event showing EMIC-driven electron precipitation. Figures \ref{fig1}(a,b,c) show measurements of hydrogen-band EMIC waves (frequency-time spectra during 06:10-06:20~UT with wave power between helium and proton gyrofrequencies) propagating quasi-parallel to the background magnetic field (wave normal angle below $30^\circ$) measured by the fluxgate magnetometer \cite{Russell16:mms} onboard the Magnetospheric Multiscale (MMS) mission \cite{Burch16}. During the observations of EMIC waves MMS mapped to the equatorial plane at $7.5 \leqslant L \leqslant 8$ and $4 \leqslant \mathrm{MLT} \leqslant 5$ \blue{\bibnote{The ranges $\Delta L = 0.5$ and $\Delta \mathrm{MLT} = 1$\,hour are obtained for the time interval of EMIC wave measurements, which are not due to the spacecraft separation, but due to a finite interval of mapping traveled by MMS.}}, a typical spatial scale of an EMIC wave source \cite{Blum17}. Here $L$ is evaluated with the empirical models \cite{Tsyganenko89,Tsyganenko95} taking into account non-dipole magnetic field configurations. There are no thermal plasma measurements during this interval. To determine plasma density, we use THEMIS-D \cite{Angelopoulos08:ssr} that crossed the same $L$ with $\Delta\mathrm{MLT}\approx 0.5$\,hour around 07:30-08:00~UT \blue{\bibnote{Note that Reference \cite{Goldstein19} in their Figure 2 showed that during the start of a very weak storm, as during this event at 6-8 UT [where $min(D_{st})$ merely reached  $-45$ nT at 12 UT] we are in a situation of plasmasphere erosion (not refilling) and plasma plumes then form only at $\sim 10$-$18$ MLT, far from the 4-5 MLT region considered in this event. Based on a $K_p$ statistical plasmapause model, the plasmapause location was then at $L < 6$ \citep{o2003empirical}. Therefore the plasma density is unlikely to have varied sensibly in one hour at $L=7.5$-$8$ and $4$-$5$ MLT. Between 6:20 UT and 7:30 UT, $D_{st}$ varied from $+13$\,nT to $-23$\,nT, and SuperMag Electrojet index increased above $300$\,nT (a substorm) after 6:20 UT. This could have led to some local changes of $B$-field at $L=7.5$. However, the region of 4-5 MLT under consideration is located rather far from the midnight region where magnetic field changes are most significant and rapid, making it unlikely that the magnetic field could have varied significantly in less than one hour during this event.}}. THEMIS-D measurements of the plasma density \cite{McFadden08:THEMIS}, further confirmed by spacecraft potential estimates \cite{Nishimura13:density}, provide the estimate of plasma frequency $\sim 15.4$\,kHz (density $\sim 3\, \mathrm{cm}^{-3}$) and the ratio of plasma frequency to electron gyrofrequency $f_{pe}/f_{ce}\approx 10$. At $\sim$6:25~UT this $L, \mathrm{MLT}$ region was crossed by the low-altitude ELFIN-B CubeSat \cite{Angelopoulos20:elfin}, as shown in Figure \ref{fig1}(g). The energetic particle detector onboard ELFIN-B measures 3D pitch-angle (angle between electron velocity and background magnetic field) electron fluxes for energies between $50$\,keV and $6$\,MeV, and thus resolves trapped (averaged over the pitch-angle range $\alpha\in(\alpha_{LC},180^\circ-\alpha_{LC})$; $\alpha_{LC}$ being the local loss-cone angle) and precipitating (averaged over $\alpha\in[0,\alpha_{LC}]$; moving sufficiently close to the Earth to be lost through collisions in the $< 100$\,km altitude ionosphere) fluxes. Figures \ref{fig1}(d,e,f) show trapped and precipitating electron fluxes, and the precipitating-to-trapped flux ratio $j_{\mathrm{prec}} / j_{\mathrm{trap}}$, respectively. Before $\sim$06:25~UT ELFIN-B crossed magnetic field lines connected to Earth's plasma sheet (i.e., trapped fluxes are mostly below $200$\,keV). There the strong magnetic field-line curvature scattering (typical for the plasma sheet \cite{Sergeev12:IB}) provides isotropic equatorial fluxes resulting in approximately equal trapped and precipitating fluxes at ELFIN-B. After $\sim$06:25~UT ELFIN-B crossed magnetic field lines connected to the outer radiation belt (i.e., energies of trapped fluxes were as high as $2$\,MeV), where the low $j_{\mathrm{prec}} / j_{\mathrm{trap}}$ is probably due to the absence of intense (whistler-mode or EMIC) resonant waves. Around 06:25:15~UT ELFIN detected strong precipitation of energetic electrons. During that time precipitating flux levels reached trapped flux levels for energies above $200$\,keV. The significant latitudinal separation of this interval from the plasma sheet excludes the field-line curvature scattering from the possible mechanisms responsible for electrons losses. The presence of $\sim 1$\,MeV precipitation, and the absence of strong $<200$\,keV precipitation exclude scattering by whistler-mode waves (which are more effective in precipitating $<100$\,keV electrons). Thus, this precipitation event is most likely driven by EMIC waves. Figure \ref{fig1}(h) shows that the peak of $j_{\mathrm{prec}} / j_{\mathrm{trap}}$ occurs at $800$\,keV, even though that ratio is high ($>0.5$) down to $400$\,keV. This is a typical example of sub-relativistic electron precipitation by EMIC waves \cite{Capannolo19,Hendry17}. 

\begin{figure}
\centering
\includegraphics[width=0.7\textwidth]{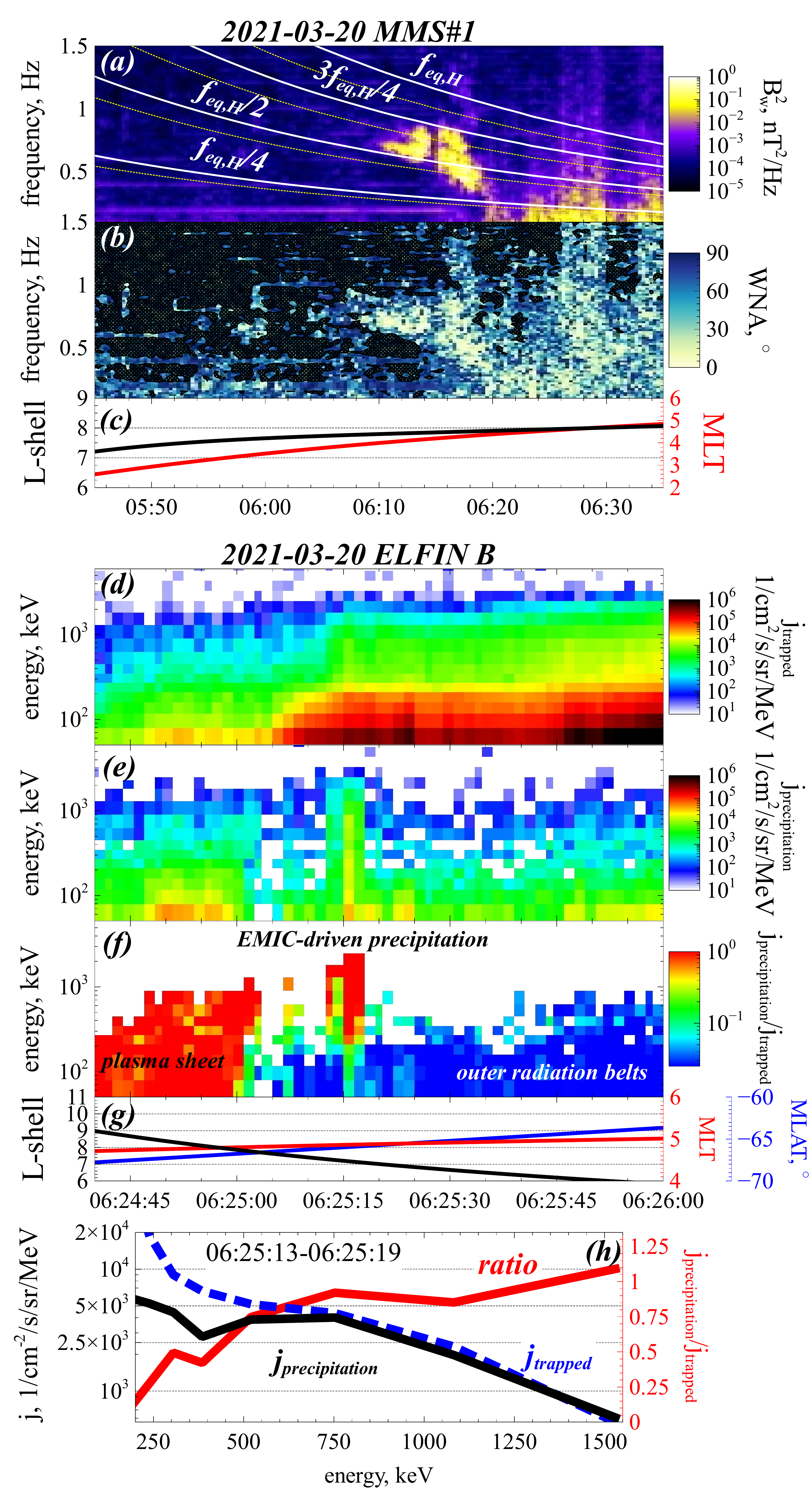}
\end{figure}
\begin{figure}
\caption{
\label{fig1} Event overview. Panels (a) and (b) show the power spectrum and wave normal angle of EMIC waves captured by MMS\,\#1, respectively. White and yellow dashed curves in panel (a) show fractions of equatorial proton gyrofrequency estimated with two models of MMS projection to the equatorial plane \cite{Tsyganenko89,Tsyganenko95}. MMS\,\#1 L-shell and $\mathrm{MLT}$ are shown in panel (c). Panels (d) and (e) show spectra of trapped and precipitating electron fluxes, respectively, measured by ELFIN-B. Panel (f) shows the precipitating-to-trapped flux ratio. ELFIN-B L-shell, magnetic latitude ($\mathrm{MLAT}$), and $\mathrm{MLT}$ are shown in panel (g). Panel (h) shows the spectrum of precipitating electrons at the moment of the strongest precipitation, 06:25:16~UT (in black), the average of trapped spectra over the precipitation event, 06:25:14-06:24:20~UT (in blue) taken as the background level of precipitation, and the ratio of the precipitating to the aforementioned background level (in red).}
\end{figure}

\begin{figure}
\centering
\includegraphics[width=0.7\textwidth]{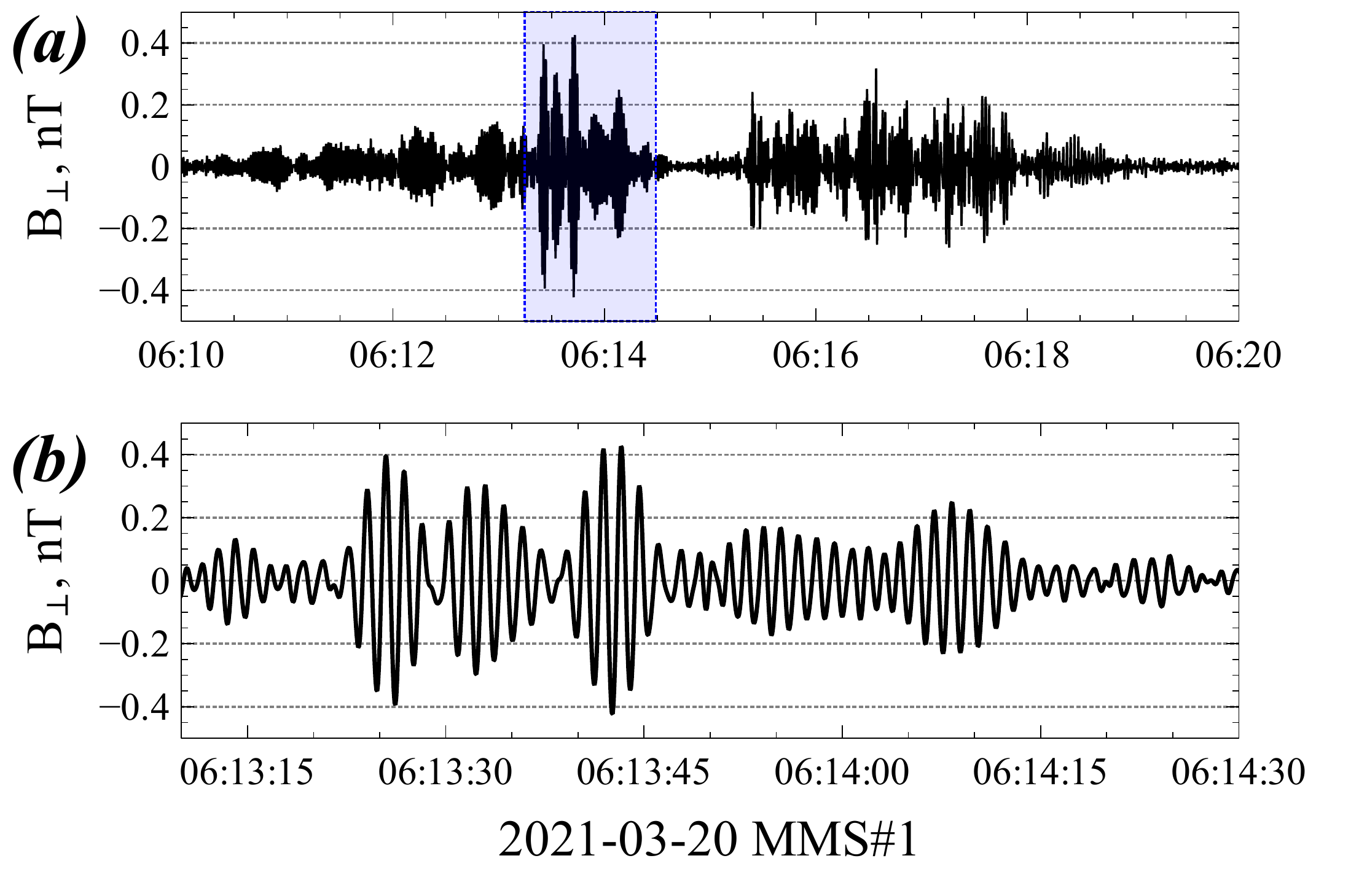}
\caption{\label{fig2}EMIC wave packets for the event from Figure \ref{fig1}. (a) The entire interval of EMIC wave observations. (b) A subinterval [shown in blue in panel (a)] with a few wave packets. The shown $B_\perp$ is orthogonal to the background magnetic field and the $x$-axis of Geocentric Solar Magnetospheric system. As EMIC wave sources can survive for hours \citep[e.g.,][]{Engebretson15,Blum20}, it is reasonable to expect that EMIC wave properties do not change significantly during the $10$-minute delay between wave measurements by MMS and electron precipitation measurements by ELFIN.}
\end{figure}

Figure \ref{fig2} shows that intense EMIC waves from the event in Figure \ref{fig1} propagate in the form of short packets, with each packet including a few wave periods. Such strong modulation of the wave field should disrupt nonlinear resonance effects (if any), causing the wave-particle resonant interaction to revert to a classical, diffusive scattering regime \cite{Tao13}. However, the sharp edges of the short wave-packets may result in nonresonant scattering \cite{Chen16:nonresonant}. Thus we shall include the wave modulation effect in the evaluation of electron diffusion rates.

The EMIC wave characteristics, plasma density, and background magnetic field give an estimate of the minimum resonant energy for electrons scattered by EMIC waves \cite{Summers&Thorne03} between $800$\,keV and $900$\,keV, depending on the magnetic field model used to project MMS to the equator. Thus, we aim to explain the observed precipitation of nonresonant electrons with energies of $400$ - $800$\,keV. Towards that goal, we generalize the quasi-linear model of electron scattering by EMIC waves. The equation of motion for an electron moving through an EMIC wave packet of wavenumber $k$ and amplitude $B_w$ propagating along the geomagnetic field line is \cite[note nonlinear terms, e.g., those included in][are not included here, because of the weak wave amplitude approximation $B_w / B \ll 1$]{Albert&Bortnik09,Grach&Demekhov20}
\begin{equation}
\begin{split}
	\frac{\mathrm{d} \varphi}{\mathrm{d} t} &= \frac{k u_z}{\gamma}  - \frac{\omega_{ce}}{\gamma} ,\\ 
	\frac{\mathrm{d} I}{\mathrm{d} t} &= \frac{u_z}{\gamma} \sqrt{\frac{2 I}{mc^2 \omega_{ce}(z)}} e B_w g(z) \sin\varphi , \\ 
\end{split}
\end{equation}
in which $t$ is time, $z$ is the field-aligned coordinate, $u_z = \gamma \mathrm{d} z / \mathrm{d} t$ is the field-aligned relativistic velocity, $\gamma$ is the Lorentz factor, $-e$ and $m$ are the charge and mass of the electron, $c$ is the speed of light, $\omega_{ce}(z) = e B(z) / (m c)$ is the electron gyrofrequency in the background magnetic field $B(z)$ given by the generalized dipole model \cite{Bell84}, $I$ is the electron magnetic moment, $\varphi$ is the phase angle between the particle perpendicular momentum and the wave magnetic field, and $g(z)$ is the shape function describing the envelope of the wave packet \cite{Bortnik&Thorne10,Bortnik15}. Because the EMIC wave phase velocity is much smaller than the electron velocity near gyroresonance, electrons are dominantly scattered in pitch angle while their energies are approximately constant, i.e., $\gamma = \sqrt{1 + (u_z/c)^2 + 2 \omega_{ce} I / (m c^2)} = \mathrm{constant}$. For weak wave intensity, the first-order change of $I$ can be obtained by integrating $\mathrm{d} I / \mathrm{d} z$ along the zeroth-order gyro-orbits \cite{morales1974effect,lamb1984behavior}
\begin{equation}
    \Delta I = \int_{z_l}^{z_u} \mathrm{d}z' \sqrt{\frac{2 I_0}{m c^2 \omega_{ce}(z')}} e B_w g(z') \sin(\varphi(z')) ,
\end{equation}
where the phase angle is
\begin{equation}
    \varphi (z) = \varphi_0 + \int_{z_l}^{z} \mathrm{d}z' \left(k(z') - \frac{\omega_{ce}(z') / c}{\sqrt{\gamma^2 - 1 - 2 \omega_{ce}(z') I_0 / (m c^2)}}\right) = \varphi_0 + \varphi_R(z) ,
\end{equation}
$z_l$ and $z_u$ are the lower and upper boundaries of the localized wave packet, respectively, $\varphi_0$ is the phase angle at the lower boundary, $\varphi_R(z)$ denotes the phase integral, and $I_0$ is the zeroth-order magnetic moment that is a constant. The variance of the magnetic moment after one pass through the localized wave packet is
\begin{equation}\label{eq:ensemble-avg-dI2}
	\left\langle \left(\Delta I\right)^2 \right\rangle = \frac{I_0 e^2 B_w^2}{m c^2 \omega_{ce}(z_c)} \left\vert \int_{z_l}^{z_u} \mathrm{d} z' g(z') e^{i \varphi_R(z')} \right\vert^2 ,
\end{equation}
where $\langle \cdot \rangle$ denotes the ensemble average over $\varphi_0$, and $z_c$ represents the center of the wave packet. We define the scattering factor
\begin{equation}\label{eq:G-exact}
	G = \left\vert \int_{z_l}^{z_u} \mathrm{d} z' g(z') e^{i \varphi_R(z')} \right\vert^2 = \left\vert \int_{-\infty}^{\infty}\mathrm{d}\kappa\,\, \hat{g}(\kappa) \int_{\Psi_l}^{\Psi_u} \frac{\mathrm{d}\Psi}{\dot{\Psi}(z)} e^{i \Psi} \right\vert^2 ,
\end{equation}
where the shape function is Fourier analyzed as $g(z) = \int_{-\infty}^{\infty} \mathrm{d}\kappa\,\, \hat{g}(\kappa) e^{i \kappa z}$, and this extra wavenumber $\kappa$ adds to the phase angle as $\Psi(z) = \kappa z + \varphi_R(z)$, $\Psi_l = \Psi(z_l)$, and $\Psi_u = \Psi(z_u)$. The most significant contribution to $G$ comes from the vicinity of the singularity $\dot{\Psi} = \mathrm{d}\Psi / \mathrm{d}z = 0$, i.e., the resonance condition. This condition leads to a solution in the real $z$ axis for resonant electrons. However, for nonresonant electrons, this condition leads to a solution of $z$ in the complex plane, and the phase integral in $\Psi$ may be evaluated by taking appropriate contours in the complex plane.

The resonance point $z_0$ is determined by the resonance condition
\begin{equation}\label{eq:resonance-condition}
    \dot{\Psi}\big\vert_{z = z_0} = \kappa + k - \frac{\omega_{ce}(z_0)}{c \sqrt{\gamma^2 - 1 - 2 \omega_{ce}(z_0) I_0 / (m c^2)}} = 0 .
\end{equation}
It is worthy to note the potential impact of the additional wavenumber $\kappa$ introduced by the shape of wave packets, especially for short packets that have relatively broad wavenumber spectra. $\kappa$ is not necessarily accounted by the EMIC wave dispersion relation, but is a formal Fourier description of the wave modulation. The details on the evaluation of $z_0$ and the mapping from $z_0$ to $\Psi_0 = \Psi(z_0)$ are given in the Appendix.

\begin{figure}[tphb]
\centering
\includegraphics[width=\textwidth]{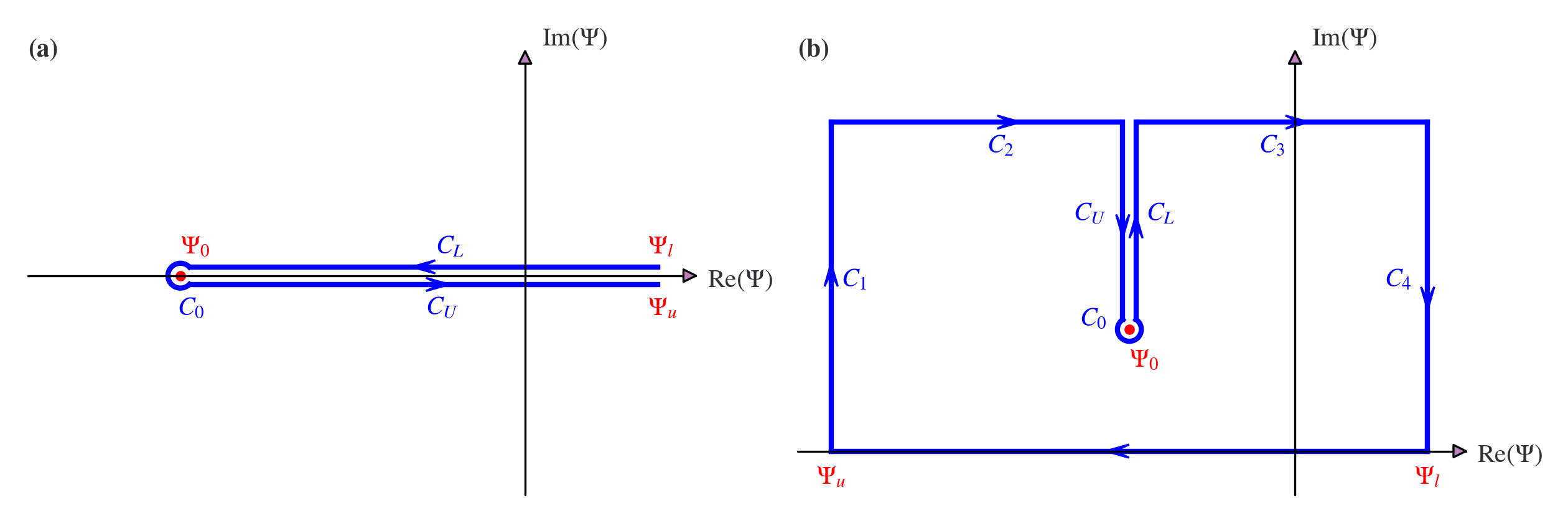}
\caption{\label{fig:contour-integral}Two regimes of resonance point $\Psi_0$ and associated contours of phase integral in $(\operatorname{Re}\Psi, \operatorname{Im}\Psi)$ for resonant and nonresonant energies. (a) Resonant regime. The resonance point $\Psi_0$ is on the real axis. The shown path of integration is called the Hankel contour \cite{krantz1999handbook}. (b) Nonresonant regime. $\Psi_0$ is in the upper half of the complex-$\Psi$ plane. The path of integration extending along the real-$\Psi$ axis from $\Psi_l$ to $\Psi_u$ is deformed into the upper half of the plane.}
\end{figure}

Since most of the contribution to scattering comes from the vicinity of the resonance point ($z_0$ or $\Psi_0$), we perform a Taylor expansion of $\Psi(z)$ and $\dot{\Psi}(z)$ around $z_0$ as: $\Psi(z) = \Psi_0 + \frac{1}{2} \ddot{\Psi}_0 \left(z - z_0\right)^2$, and $\dot{\Psi}(z) = \ddot{\Psi}_0 \left(z - z_0\right)$, where $\ddot{\Psi}_0 = \mathrm{d}^2 \Psi(z_0) / \mathrm{d} z^2$. Using these expansions, we express $\dot{\Psi}$ in terms of $\Psi$: $\dot{\Psi} = \left[2 \ddot{\Psi}_0 \left(\Psi - \Psi_0\right)\right]^{\frac{1}{2}}$. Thus the phase integral is written as
\begin{equation}\label{eq:phase-integral}
	\int_{\Psi_l}^{\Psi_u} \frac{\mathrm{d}\Psi}{\dot{\Psi}(z)} e^{i \Psi} = \frac{1}{(2 \ddot{\Psi}_0)^{\frac{1}{2}}} e^{i \Psi_0} 	\int_{\Psi_l}^{\Psi_u} \frac{\mathrm{d}\Psi}{(\Psi - \Psi_0)^{\frac{1}{2}}} e^{i (\Psi - \Psi_0)} .
\end{equation}
Because the denominator in the integrand vanishes at $\Psi = \Psi_0$ and appears as a fractional power of $\Psi - \Psi_0$, $\Psi_0$ is a branch point such that the contour of the integral must go around this point through an infinitesimally small circuit. Such contours are shown in Figures \ref{fig:contour-integral}(a) and \ref{fig:contour-integral}(b) for resonant and nonresonant energies, respectively. Interestingly, the underlying structure of the contour integral for nonresonant energies is similar to that of field-line curvature scattering \cite{Cohen78,Birmingham84,Xu&Egedal22}. It turns out that, for either resonant or nonresonant regimes, the phase integral only survives on the branch cuts labeled $C_L$ and $C_U$. In the latter regime, using the Cauchy integral theorem, the phase integral can be evaluated as
\begin{equation}
    \int_{\Psi_l}^{\Psi_u} \frac{\mathrm{d}\Psi}{\dot{\Psi}(z)} e^{i \Psi} = -\frac{e^{i \Psi_0}}{(2 \ddot{\Psi}_0)^{\frac{1}{2}}} \left(\int_{C_U} + \int_{C_L}\right) \frac{\mathrm{d}\Psi}{(\Psi - \Psi_0)^{\frac{1}{2}}} e^{i (\Psi - \Psi_0)} = \left(\frac{2 \pi}{\ddot{\Psi}_0}\right)^{\frac{1}{2}} e^{i (\Psi_0 + \frac{\pi}{4})} ,
\end{equation}
which is the same as that for resonant scattering except for a trivial phase factor $e^{i \pi}$. 

Thus the scattering factor for both resonant and nonresonant regimes can be expressed in one unified formula:
\begin{equation}\label{eq:G-complex}
	G = 2 \pi \left\vert \int_{-\infty}^{\infty}\mathrm{d}\kappa\,\, \hat{g}(\kappa) \frac{e^{i \Psi_0}}{(\ddot{\Psi}_0)^{\frac{1}{2}}} \right\vert^2 .
\end{equation}
In the limit of an infinitely long wave packet $g(z) = 1$ corresponding to $\hat{g}(\kappa) = \delta(\kappa)$, we have
\begin{equation}\label{eq:G-infinite}
	G = 2 \pi \frac{e^{- 2 \operatorname{Im}(\Psi_0)}}{\vert \ddot{\Psi}_0 \vert} .
\end{equation}
This ``infinite packet'' limit shows that the exponential decay of the scattering efficacy away from resonance is controlled by the imaginary part of the resonance point in the complex-$\Psi$ plane. This decay rate converges to $0$ for the resonant regime. In addition, the denominator $\ddot{\Psi}_0 \propto \mathrm{d}\omega_{ce}(z_0) / \mathrm{d}z$ recovers the dependence of the resonant scattering rate on the magnetic field inhomogeneity in the narrowband limit \cite{Albert10}.

The bounce-averaged pitch angle diffusion rate is defined as $D_{\alpha\alpha} = \langle(\Delta \alpha_{\mathrm{eq}})^2\rangle / (2 \tau_b)$, where $\langle(\Delta \alpha_{\mathrm{eq}})^2\rangle$ is the variance of equatorial pitch angle caused by wave packets in one bounce period $\tau_b$. Using $\sin^2{\alpha_{\mathrm{eq}}} = 2 I \omega_{ce, \mathrm{eq}}/ (\gamma^2 - 1) m c^2$, Equations \eqref{eq:ensemble-avg-dI2} and \eqref{eq:G-exact}, we obtain a generalized pitch angle diffusion rate for both resonant and nonresonant regimes
\begin{equation}\label{eq:Daa}
    D_{\alpha\alpha} = \frac{e^2 B_w^2 G \omega_{ce, \mathrm{eq}}}{2 (\gamma^2 - 1) m^2 c^4 \cos^2{\alpha_{\mathrm{eq}}} \omega_{ce}(z_c) \tau_b} ,
\end{equation}
where the information about the wave shape function and resonant/nonresonant regimes is embedded in the scattering factor $G$.

To verify our theoretical prediction of the generalized pitch angle diffusion rate and demonstrate the effect of wave packet size, we perform a series of test particle simulations for a realistic example. Following spacecraft observations in Figure \ref{fig1}, we use an equatorial geomagnetic field $B_{\mathrm{eq}} \approx 55$\,nT and the estimated plasma density $3\,\mathrm{cm}^{-3}$. The dipole magnetic field can be approximated as $B_{0z} = B_{\mathrm{eq}} (1 + \xi z^2)$ around the equator $z = 0$ \cite{Bell84}. Because the EMIC wave packet is essentially static as seen by relativistic electrons, the wave magnetic fields are $\delta B_x = B_w e^{-z^2/(2L_z^2)} \cos{(kz)}$,  $\delta B_y = B_w e^{-z^2/(2L_z^2)} \sin{(kz)}$, and $\delta B_z = 0$, where the wavenumber is $k = 1.79\, d_i^{-1}$ ($d_i$ being the ion inertial length) from the cold plasma dispersion relation, the wave amplitude is $B_w = 10^{-3} B_{\mathrm{eq}}$, and $L_z$ is the characteristic packet size. Based on these parameters, the minimum electron resonant energy is $0.83$\,MeV for $\alpha_{\mathrm{eq}} = 10^{\circ}$. We initialize an ensemble of $10^6$ electrons uniformly distributed in gyrophase, all of which have the same initial equatorial pitch angle $\alpha_{\mathrm{eq}}$ (fixed at $10^\circ$) and the same initial kinetic energy $E_k$ (scanned from $0.1$ to $0.83$\,MeV). These electrons are launched at a location in the southern hemisphere well outside the wave packet and moving in the $+z$ direction. We collect ensemble electron statistics on the other side of the wave packet and calculate the pitch angle scattering rate.

\begin{figure}[tphb]
\centering
\includegraphics[width=\textwidth]{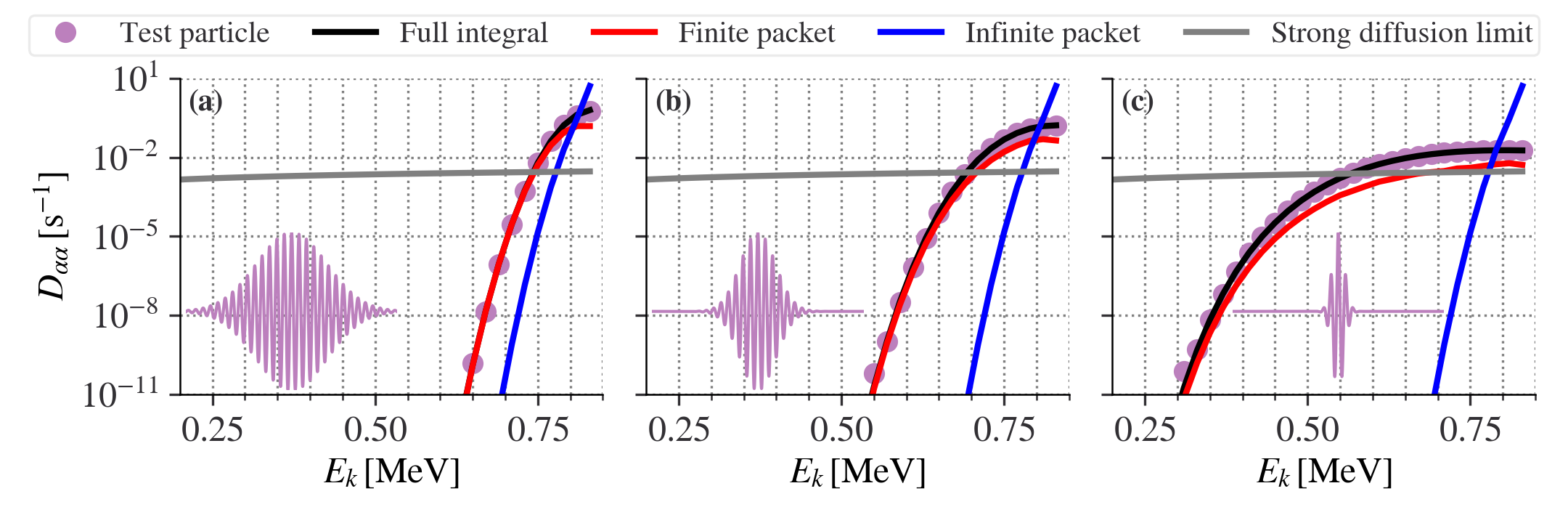}
\caption{\label{fig:theory-simulation-comparison}Comparison of pitch angle diffusion rate as a function of energy between theory and simulations for three different packet sizes. (a) $k L_z = 30$. (b) $k L_z = 15$. (c) $k L_z = 5$. The inset plots sketch the EMIC wave-packets. Test particle simulation results are shown in magenta dots. Theoretical predictions of $D_{\alpha\alpha}$ using scattering factors from Equations \eqref{eq:G-exact}, \eqref{eq:G-complex}, and \eqref{eq:G-infinite} are shown in black, red, and blue lines, respectively. The diffusion rates from the simulations with $B_w / B_{\mathrm{eq}} = 10^{-3}$ are scaled up to $B_w / B_{\mathrm{eq}} = 10^{-2}$ (as seen in the spacecraft observations) to compare them directly with those in the strong diffusion limit, depicted by the grey lines. The lower wave amplitude used in simulations is intended to separate nonresonant from nonlinear effects and to be more consistent with the perturbation analysis.}
\end{figure}

Figure \ref{fig:theory-simulation-comparison}(a) shows the results for a relatively long wave packet $k L_z = 30$. For comparison, the predicted pitch angle diffusion rate from Equation \eqref{eq:Daa} is calculated using three versions of the scattering factor $G$: Equation \eqref{eq:G-exact} of the full integral, Equation \eqref{eq:G-complex} of the finite packet, and Equation \eqref{eq:G-infinite} in the limit of an infinite packet. The finite packet results capture the exponential decay of $D_{\alpha\alpha}$ below the minimum resonant energy. Compared to test particle simulations and full integrals, evaluating $D_{\alpha\alpha}$ using Equation \eqref{eq:G-complex} is computationally more efficient (because the phase integral has been carried out analytically) at the expense of sacrificing a small degree of accuracy (caused by the Taylor expansion around the resonance point). As we continue decreasing the packet size to $k L_z = 15$ [Figure \ref{fig:theory-simulation-comparison}(b)] and further to $k L_z = 5$ [Figure \ref{fig:theory-simulation-comparison}(c)], the energy range having significant pitch angle diffusion rates is greatly extended. Notably, EMIC waves with approximately $5$ wave periods within a single packet can extend electron scattering energy from $830$\,keV to $550$\,keV without significantly reducing the pitch angle diffusion rate [Figure \ref{fig:theory-simulation-comparison}(c)]. Compared to the strong diffusion limit \cite[associated with $j_{\mathrm{prec}}/j_{\mathrm{trap}} \sim 1$, see][]{Kennel69}, realistically short wave-packets $k L_z = 5$ [Figure \ref{fig2}(b)] provide strong scattering of electrons well below the minimum resonant energy, and thus can explain the observed precipitation of $400$ - $800$ keV electrons [Figure \ref{fig1}(h)]. Because the spread of the shape function in wavenumber space, $\hat{g}(\kappa) \propto e^{-\kappa^2 L_z^2 / 2}$, is of the order of $1 / L_z$, shorter wave packets associated with wider wavenumber spectra can increase the original wavenumber more effectively, to $k + (1 / L_z)$, and thus lead to an equivalent ``resonant'' scattering even below the minimum resonant energy for the original wavenumber $k$. This demonstration, together with the observation of short packets in equatorial spacecraft data simultaneous with the electron precipitation, confirms our hypothesis that nonresonant scattering of relativistic electrons by EMIC waves below the minimum resonant energy can be a significant contributor to the overall EMIC contribution to relativistic electron losses. 

Note that $D_{\alpha\alpha} \approx 0.01 / \mathrm{s}$ at $550$\,keV provides an equilibrium level of $j_{\mathrm{prec}}/j_{\mathrm{trap}} \approx 0.55$ caused by diffusive scattering by the observed short packets \cite[see][for the equation relating $j_{\mathrm{prec}}/j_{\mathrm{trap}}$ with $D_{\alpha\alpha}$]{Kennel&Petschek66,Li13:POES}, sufficient to account for the observed precipitating fluxes [Figure \ref{fig1}(h)] after a few seconds. However, the strongest $j_{\mathrm{prec}}/j_{\mathrm{trap}} \approx 1$ at resonant energies should be caused by non-diffusive scattering \cite[e.g.,][]{Kubota&Omura17,Nakamura19:emic} by intense EMIC waves \citep[e.g., rising tone EMICs; see][]{Nakamura14:emic}.

In summary, we have generalized the quasi-linear diffusion model to include nonresonant wave-particle interactions. The diffusion rate exponentially decays away from the resonance, where the decay rate is controlled by the imaginary part of the resonance point in the complex phase plane. Using this generalized formulation of the interaction for arbitrary EMIC waveforms, we have demonstrated that short EMIC wave-packets greatly enhance the scattering rate for sub-relativistic electrons by introducing an appreciable spread of wave power in wavenumber space, and can account for the precipitation of these electrons observed at low-Earth orbit. Our approach of including nonresonant wave-particle interactions can be readily used in radiation belt modelling, and more broadly, in other plasma systems where sufficiently strong and short wave-packets render nonresonant effects important.

\begin{acknowledgments}
The work of X.A., A.A., V.A., X.Z., and J.B. was supported by NSF awards 2108582 and 2021749, and NASA awards 80NSSC20K1270 and 80NSSC20K0917.  ELFIN data is available at \url{https://data.elfin.ucla.edu/} and online summary plots at \url{https://plots.elfin.ucla.edu/summary.php}. THEMIS data is available at \url{http://themis.ssl.berkeley.edu/}. We acknowledge MMS FGM data obtained from \url{https://lasp.colorado.edu/mms}. Data access and processing was done using SPEDAS V4.1 \cite{Angelopoulos19}. We would like to acknowledge high-performance computing support from Cheyenne (doi:10.5065/D6RX99HX) provided by NCAR's Computational and Information Systems Laboratory, sponsored by the National Science Foundation \cite{cheyenne}.
\end{acknowledgments}

\textbf{Appendix on the evaluation of $z_0$ and the mapping from $z_0$ to $\Psi_0$.}--- Using a reduced dipole model around the equator $\omega_{ce} (z) = \omega_{ce, \mathrm{eq}} (1 + \xi z^2)$ ($\omega_{ce, \mathrm{eq}}$ being the equatorial electron gyrofrequency), we solve Equation \eqref{eq:resonance-condition} to obtain
\begin{equation}
	z_0 = 
	\begin{cases}
		\pm \sqrt{\left(\frac{\omega_{ce, R}}{\omega_{ce, \mathrm{eq}}} - 1\right) / \xi} & \text{for } \omega_{ce,R} \geqslant \omega_{ce, \mathrm{eq}} \text{ (resonant)}  , \\
		\pm i \sqrt{\left(1 - \frac{\omega_{ce, R}}{\omega_{ce, \mathrm{eq}}}\right) / \xi} & \text{for } \omega_{ce,R} < \omega_{ce, \mathrm{eq}} \text{ (nonresonant)} ,
	\end{cases}
\end{equation}
where
\begin{equation}
    \omega_{ce, R} = (\kappa + k)^2 c^2 \left[-(I_0 / m c^2) + \sqrt{(I_0 / m c^2)^2 + (\gamma^2 - 1) (\kappa + k)^{-2} c^{-2}}\right]
\end{equation}
and $\xi = 9 / (2 L^2 R_E^2)$ ($R_E$ being the Earth radius). If electrons have sufficiently high energies such that $\omega_{ce,R} \geqslant \omega_{ce, \mathrm{eq}}$, resonance occurs at the equator or at higher latitudes. Below the resonant energies (i.e., $\omega_{ce,R} < \omega_{ce, \mathrm{eq}}$), we can still get a resonance point but now located in the complex plane. It is convenient to map $z_0$ to $\Psi_0 = \Psi(z_0)$. Such mapping yields $\operatorname{Im}(\Psi_0) = 0$ for resonant energies, whereas for nonresonant energies, it gives an imaginary part in the upper half of the complex plane:
\begin{equation}
\begin{split}
    \operatorname{Im}(\Psi_0) &= \frac{1}{2 \sqrt{\xi}} \left\{ -\left(1 - \frac{\omega_{ce, R}}{\omega_{ce, \mathrm{eq}}}\right)^{1/2} \left[\frac{3 (\kappa + k)}{2} + \frac{(\kappa + k)}{2}\left(1 + \frac{\gamma^2 - 1}{(\kappa + k)^2 (I_0/mc)^2}\right)^{1/2}\right] \right. \\
		&\left. + \frac{(\gamma^2 -1) m c^2 + 2 I_0 \omega_{ce, \mathrm{eq}}}{(2 I_0)^{3/2} (\omega_{ce, \mathrm{eq}} / m)^{1/2}} \ln\left[\frac{\left(\frac{(\gamma^2 - 1) m c^2}{2 I_0 \omega_{ce, \mathrm{eq}}} - 1\right)^{1/2}}{\left(\frac{(\gamma^2 - 1) m c^2}{2 I_0 \omega_{ce, \mathrm{eq}}} - \frac{\omega_{ce, R}}{\omega_{ce, \mathrm{eq}}}\right)^{1/2} - \left(1 - \frac{\omega_{ce, R}}{\omega_{ce, \mathrm{eq}}}\right)^{1/2}}\right] \right\} .
\end{split}
\end{equation}
The imaginary part of $\Psi_0$ distinguishes nonresonant from resonant scattering. The real part of $\Psi_0$ is not important here.


\begin{thebibliography}{72}
\bibitem [{1()}]{1}%
	\BibitemOpen
	\href@noop {} {}\bibinfo {note} {The ranges $\Delta L = 0.5$ and $\Delta
		\mathrm {MLT} = 1$\,hour are obtained for the time interval of EMIC wave
		measurements, which are not due to the spacecraft separation, but due to a
		finite interval of mapping traveled by MMS.}\BibitemShut {Stop}%
\end{thebibliography}

%

\end{document}